\documentclass[aps,preprint]{revtex4}

\usepackage{graphicx}

\newcommand{\be}{\begin{equation}}
\newcommand{\ee}{\end{equation}}

\begin{document}

\title{\bf Lowest Order Constrained Variational Calculation
 of Structure Properties of Protoneutron Star}
\author{G.H. Bordbar\footnote{Corresponding author},
 S.M. Zebarjad and R. Zahedinia}
 \affiliation{
Department of Physics, Shiraz University,
Shiraz 71454, Iran\footnote{Permanent address}\\
and\\
Research Institute for Astronomy and Astrophysics of Maragha,\\
P.O. Box 55134-441, Maragha, Iran }
\begin{abstract}
We calculate the structure properties of protoneutron star such as
equation of state, maximum mass, radius and temperature profile
using the lowest order constrained variational method. We show
that the mass and radius of protoneutron star decrease by
decreasing both entropy and temperature. For the protoneutron
star, it is shown that the temperature is nearly constant in the
core and drops rapidly near the crust.
\end{abstract}

\maketitle

\newpage
\section{Introduction}
Neutron star which is highly compact stellar objects is a result
of supernova explosion. The structure properties of this object,
especially its maximum mass, is of a great interest for
astrophysicists.
 Since the compactness parameter for a neutron star is about $0.2-0.4$,
 its structure should be studied using the general
theory of relativity. In fact, the computation of the structure
properties of a neutron star can be derived using the
Tolman-Oppenheimer-Volkoff (TOV) equation \cite{1}.

Just after the supernova collapse, a newly-born neutron star
called protoneutron star is formed.  At these stages, the neutron
stars are rich in leptons. This is due to the fact that the
neutrinos are trapped in the protoneutron star matter. The
temperature of protoneutron star is greater than $10 MeV$
\cite{2,3,4,5,6,7,8,9}. Therefore, the high temperature of these
stages cannot be neglected with respect to the Fermi temperature
throughout the calculation of its structure. Depending on the
total number of nucleons, a protoneutron star evolves either to
black hole or to stable neutron star \cite{3,7,8,9}. Therefore,
computation of the maximum mass of protoneutron star is of crucial
importance.

In recent years, we have investigated the properties of neutron
star matter \cite{10,11,12,13,14}. In the present work, we intend
to compute the structure properties of protoneutron star at
different stages employing the lowest order constrained
variational approach using the modern $AV_{18}$ potential
\cite{wiringa}.
\section{Formalism}
In this section, we give the formalism of calculation for the
protoneutron star matter equation of state.

We write the total energy per baryon ($E$) as the sum of
contributions from leptons and nucleons,
\begin{equation}
E = E_{lep} + E_{nucl}\cdot
\end{equation}
The contribution from the energy of leptons is
\begin{equation}
E_{lep} = \frac{m^4c^5}{\pi^2\rho\hbar^3}\int_0^\infty n(x)\sqrt{1
+ x^2} x^2 dx, \label{Elep}
\end{equation}
where $\rho$ is the total number density of nucleons and $n(x)$ is
the Fermi-Dirac distribution function,  where $x$ defined,
\begin{equation}
x = \frac{\hbar k}{m c}\cdot
\end{equation}
We calculate the contribution from the energy of nucleons using
the lowest order constrained variational method. We consider up to
the two-body term in the cluster expansion for the energy
functional,
\begin{equation}
E_{nucl} = E_1 + E_2\cdot
\end{equation}
The one-body energy $E_1$ is
\begin{equation}
E_1 = \sum_{i=n,p} \frac{\hbar^2}{2m_i\rho\pi^2} \int_0^\infty
n_i(k) k^4 dk\cdot
\end{equation}
The two-body energy $E_2$ is
\begin{equation}
E_2 = (2A)^{-1} \sum_{ij} <ij\mid \{-\frac{\hbar^2}{2m} \left[
f(12), [\nabla_{12}, f(12)]\right] + f(12)V(12)f(12)\}\mid ij>_a,
\end{equation}
where f(12) and V(12) are the two-body correlation function and
nucleon-nucleon potential, respectively. V(12) has the following
general form \cite{wiringa},
\begin{equation}
V(12) = \sum_{p=1}^{18} V^p(r_{12}) O^p_{12}\cdot
\end{equation}
Finally, we minimize the two-body energy, $E_2$, with respect to
the variation in the two-body correlation function subjected to
the normalization constraint. From this minimization, we get a set
of Euler-Lagrange differential equations. The correlation
functions are calculated by solving these differential equations
and then, the two-body energy, $E_2$, is computed \cite{17,18}.
This leads to the equations of state used in our present work.

\section{Calculations of Protoneutron Star Structure Properties}
In this section, we calculate the structure properties of neutron
star in the different stages just after its formation by
numerically integrating the TOV equation. The structure properties
of neutron star in these stages such as maximum mass, radius and
temperature profile are calculated using the modern microscopic
equations of state derived from constrained variational method
employing the Argonne $V_{18}$ potential \cite{wiringa} as the
inter-nucleonic interaction. At low densities, we also consider a
hot dense matter model and use the equations of state obtained by
Gondek et al. \cite{15} and Strobel et al. \cite{16}. Here is our
results for different stages.

\subsection{Lepton rich protoneutron star}

Since a neutron star, at the beginning of its lifetime
(protoneutron star), is opaque with respect to neutrinos
therefore, it contains a high lepton fraction ($Y$), $0.3-0.4$.
Furthermore, at this stage of neutron star, the entropy per baryon
($s$) is nearly constant throughout the star, $1-2 k_B $. Since,
we are also dealing with uncharged neutron star matter, the lepton
and proton fractions should be equal \cite{2,3,4,5}.

Our results for the pressure of protoneutron star matter as a
function of mass density are presented in Fig. 1 at entropies $s=1
k_B$ and $s=2 k_B$ with $Y=0.4$ and $Y=0.3$. It is seen that for a
fixed value of entropy, the pressure of protoneutron star matter
increases by decreasing the lepton fraction. We see that for a
given value of lepton fraction, the equation of state for
$s=2.0k_B$ is stiffer than for $s=1.0k_B$. In Fig. 1, We have
compared our results with those of Gondek et al. \cite{15} and
Strobel et al. \cite{16}. We have seen that the results of Gondek
et al. \cite{15} are nearly in agreement with our results for
$s=2.0 k_B$ and $Y=0.3$. But, the equations of state calculated by
Strobel et al. \cite{16} are stiffer than those of ours.

In Fig. 2, we have presented the gravitational mass of
protoneutron star as a function of central mass density for
different values of entropy and lepton fraction. From Fig. 2, we
can see that the gravitational mass increases by increasing the
central mass density. It is shown that for different cases of
entropy and lepton fraction, the gravitational mass exhibits
different  mass limits (maximum mass) as given in Table 1. From
this table, the decreasing of maximum mass by decreasing the
entropy and increasing the lepton fraction is seen. In Fig. 2 and
Table 1, the results of Gondek et al. \cite{15} and Strobel et al.
\cite{16} are also given for comparison. It is seen that these
results are different from our calculations.

The radius of protoneutron star as a function of central mass
density is shown in  Fig. 3 for different entropies and lepton
fractions. It is seen that as the central mass density increases,
the radius decreases very rapidly and then reaches a nearly
constant value. The radius of protoneutron star corresponding  to
the maximum mass for different entropies and lepton fractions is
given in Table 2. It can be seen that the stiffer equation of
state leads to the relatively higher radius.

The variations of temperature throughout the protoneutron star
matter with respect to the radial coordinate ($r$) are presented
in Figs. 4 and 5 for $s=1 k_B$ and $s=2 k_B$ with $Y=0.4$ and
$Y=0.3$. We can see from the Figs. 4 and 5 that the temperature is
decreasing slowly by increasing $r$, but it drops rapidly near the
protoneutron star crust.

\subsection{Beta-stable protoneutron star}

After  complete deleptonization, neutrino trapped within the hot
interior matter of neutron star do not affect the beta stability
condition and therefore the lepton fraction is determined from the
beta-equilibrium criteria \cite{2,3,4,5}. In this stage, we have
calculated the structure properties of protoneutron star for both
isentropic and isothermal paths.

\subsubsection{Isentropic paths}

For different values of entropy, our calculated equations of state
of beta-stable protoneutron star are shown in Fig. 6. It is seen
that the pressure of beta-stable protoneutron star matter
increases by increasing entropy. In Fig. 6, the results of Strobel
et al. \cite{16} are also plotted for comparison. There is a
compatible difference between our results and those of Strobel et
al. \cite{16}.

Our calculated gravitational mass of beta-stable protoneutron star
as a function of central mass density for $s=1.0k_B$ and
$s=2.0k_B$ is presented in Fig. 7. From this figure, the
increasing of gravitational mass by increasing both central mass
density and entropy is seen. As it is shown in Fig. 7, for higher
values of central mass density, the gravitational mass shows a
limiting value (maximum mass). The maximum mass of beta-stable
protoneutron star is given in Table 3 for different entropies. It
is seen that the maximum mass decreases by decreasing the entropy.
In Fig. 7 and Table 3, we have compared our results with the
results of Strobel et al. \cite{16}. It is seen that our results
are different from those of Strobel et al. \cite{16}.

In Fig. 8, our results for the radius of beta-stable protoneutron
star as a function of central mass density are plotted at
different values of entropy. We see that the radius decreases very
rapidly by increasing the central mass density and then exhibits a
nearly constant value. The radius corresponding to the maximum
mass of beta stable protoneutron star for $s=1.0k_B$ and
$s=2.0k_B$ are also given in Table 3. This shows the decreasing of
radius with respect to decreasing entropy.

The temperature of beta-stable protoneutron star matter as a
function of radial coordinate ($r$) is shown in Fig. 9 for
different entropies. The same as Fig.s 4 and 5, we see that the
temperature decreases slowly by increasing $r$, but drops rapidly
near the beta-stable protoneutron star crust.

\subsubsection{Isothermal paths}
We are now able to do  the above calculations for the protoneutron
star in the beta-equilibrium case at different isothermal paths.

Our results for the equation of state, gravitational mass and
radius of beta-stable protoneutron star at different values of
temperature are presented in Figs. 10-12, respectively. At
different temperatures, we have extracted the maximum mass and
corresponding radius from these figures. Our results are given in
Table 4. From this table, we can see that the maximum mass and
radius of beta-stable protoneutron star decrease by decreasing the
temperature.

\section{Summary and Conclusion}
The protoneutron stars  which are lepton rich and hot objects are
formed just after the supernova explosion. In
this paper, we have integrated TOV equation to compute
the maximum mass, radius and temperature profile for this
object. As we have shown at a fixed value of lepton fraction,
the mass and radius of protoneutron star
decreases by decreasing entropy. The temperature is nearly constant
in the core of protoneutron star and drops
rapidly near its crust. The same properties of protoneutron star
at beta equilibrium condition have been also
calculated along the isentropic and isothermal paths.
We have shown that the  maximum mass and corresponding
radius  increase by increasing temperature.
\acknowledgements{ This work has been supported by Research
Institute for Astronomy and Astrophysics of Maragha. We wish to
thanks  Shiraz University Research Council.}

\newpage
\begin{table}
\begin{center}
\caption{Our results for the maximum mass ($M_{\odot}$) of
protoneutron star at different values of entropy ($s$) and lepton
fraction ($Y$). The results of Gondek et al. [18] (GHZ) and
Strobel et al. [19] (SSW) are also given for comparison.}
\label{tab1}
\begin{tabular}{|c|c|c|}
\hline
&$s=1.0 k_B$&$s=2.0 k_B$\\
\hline
$Y=0.4$&1.55&1.56\\
$Y=0.3$&1.63&1.70\\
SSW, $Y=0.4$&2.01&2.03\\
GHZ, $Y=0.4$&&1.91\\ \hline
\end{tabular}
\end{center}
\end{table}
\begin{table}
\begin{center}
\caption{Our results for the radius ($km$) of protoneutron star
corresponding to the maximum mass at different entropies and
lepton fractions.} \label{tab2}
\begin{tabular}{|c|c|c|}
\hline
&$s=1.0 k_B$&$s=2.0 k_B$\\
\hline
$Y=0.4$&7.62&7.68\\
$Y=0.3$&8.13&8.35\\
\hline
\end{tabular}
\end{center}
\end{table}
\begin{table}
\begin{center}
\caption{Our results for the maximum mass ($M_{\odot}$ ) of
beta-stable protoneutron star and corresponding radius ($km$) at
different entropies. The results of Strobel et al. [19] (SSW) for
the maximum mass of beta-stable protoneutron star are also given
for comparison.} \label{tab3}
\begin{tabular}{|c|c|c|}
\hline
&$s=1.0 k_B$&$s=2.0 k_B$\\
\hline
Mass&1.68&1.76\\
Radius&8.05&8.15\\
SSW&1.98&2.03\\
\hline
\end{tabular}
\end{center}
\end{table}
\begin{table}
\begin{center}
\caption{Our results for the maximum mass ($M_{\odot}$ ) of
beta-stable protoneutron star and corresponding radius ($km$) at
different temperatures. } \label{tab4}
\begin{tabular}{|c|c|c|c|}
\hline
&$T=5.0 MeV$&$T=10.0 MeV$& $T=20.0 MeV$\\
\hline
Mass&2.19&2.20&2.23\\
Radius&7.85&7.88&8.01\\
\hline
\end{tabular}
\end{center}
\end{table}

\newpage
\begin{figure}
 \includegraphics{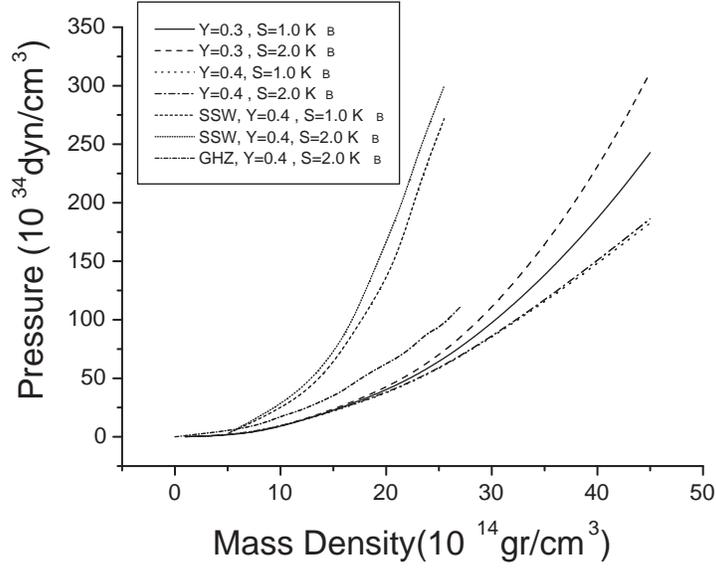}
 \caption{Our results for the pressure ($10^{34}dyn/cm^2$) of the protoneutron
star matter versus mass density ($10^{14} gr/cm^3$) at entropies
$s=1 k_B$ and $s=2 k_B$ with lepton fractions $Y=0.4$ and $0.3$.
The results of Gondek et al. [18] (GHZ) and Strobel et al. [19]
(SSW) are also given for comparison.} \label{fig1}
\end{figure}
\newpage
\begin{figure}
 \includegraphics{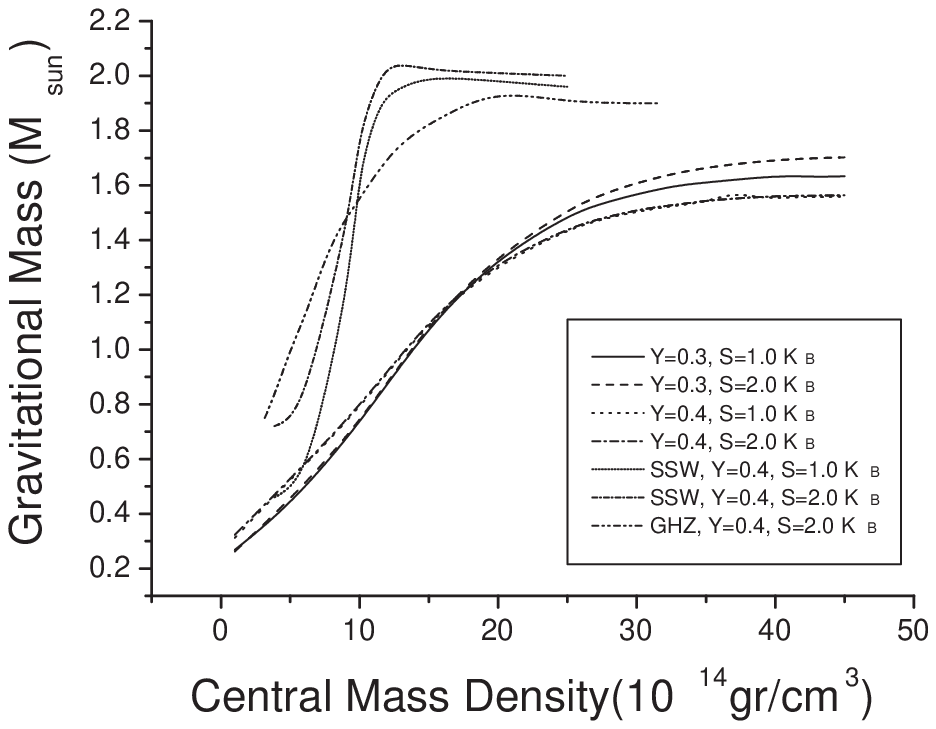}
 \caption{As Fig. 1, but for the gravitational mass ($M_{\odot}$) of protoneutron
 star versus central mass density ($10^{14} gr/cm^3$). }
 \label{fig2}
\end{figure}

\newpage
\begin{figure}
 \includegraphics{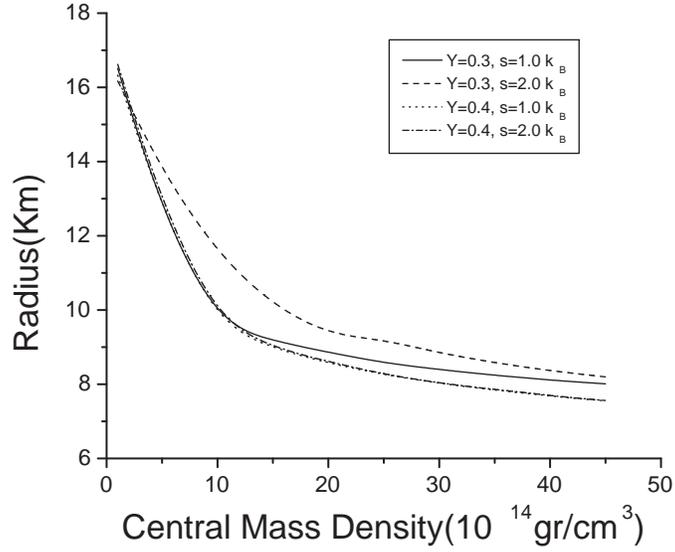}
 \caption{Our results for the radius ($km$) of protoneutron star
 versus central mass density ($10^{14} gr/cm^3$) at entropies $s=1 k_B$
 and $s=2 k_B$ with lepton fractions $Y=0.4$ and $0.3$. }
  \label{fig3}
\end{figure}
\newpage
\begin{figure}
 \includegraphics{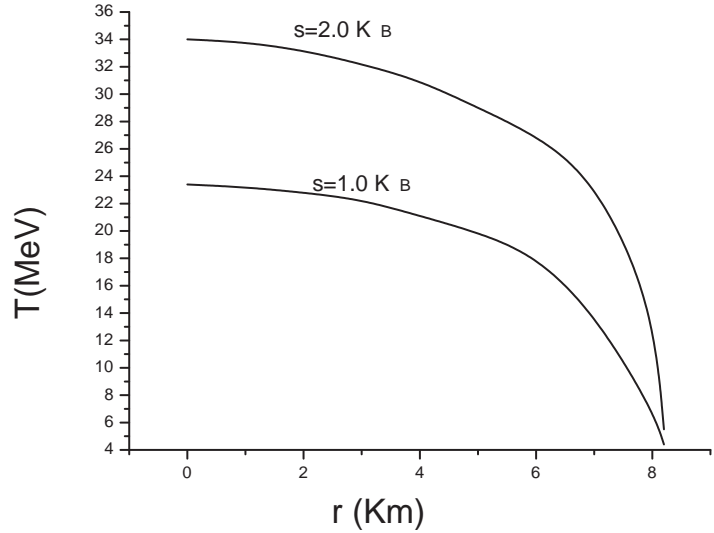}
 \caption{Our results for the temperature profile  of
  protoneutron star at different entropies for $Y=0.4$.
  }
  \label{fig4}
  \end{figure}
\newpage
\begin{figure}
 \includegraphics{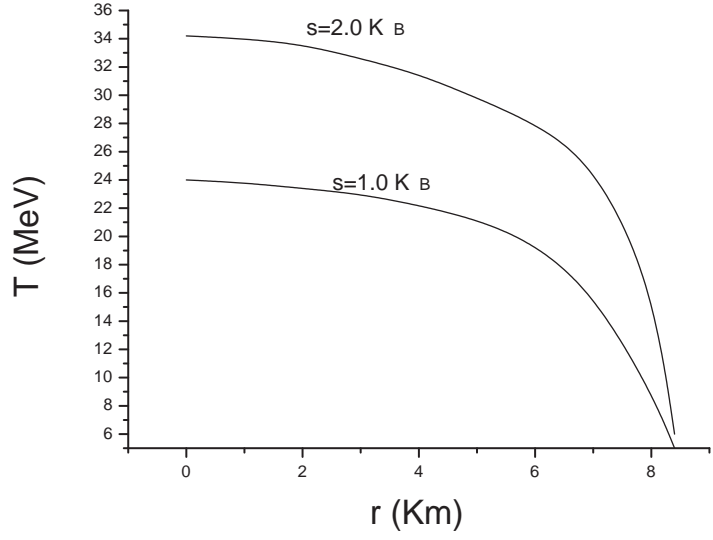}
 \caption{As Fig. 4, but for $Y=0.3$.
  }
  \label{fig5}
  \end{figure}
\newpage
\begin{figure}
 \includegraphics{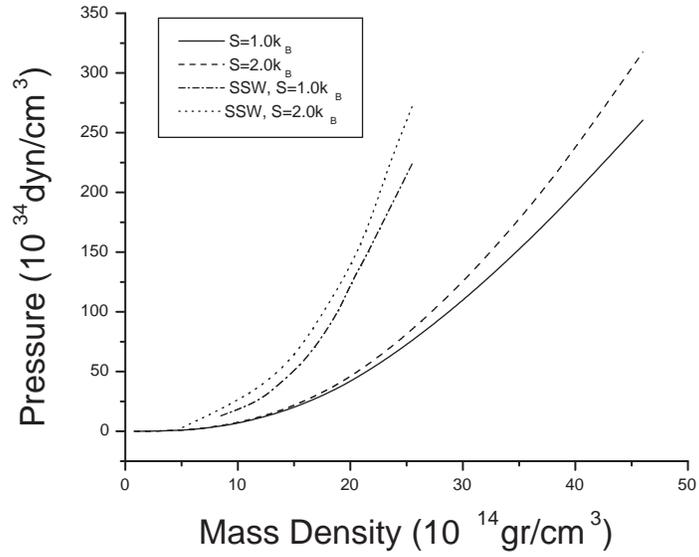}
 \caption{Our results for the pressure ($10^{34} dyn/cm^2$) of
 beta-stable protoneutron star matter as a
 function of mass density ($10^{14} gr/cm^3$) at different entropies.
 The results of Strobel et al. [19] (SSW) are also given for
 comparison.
  }
  \label{fig6}
\end{figure}
\newpage
\begin{figure}
 \includegraphics{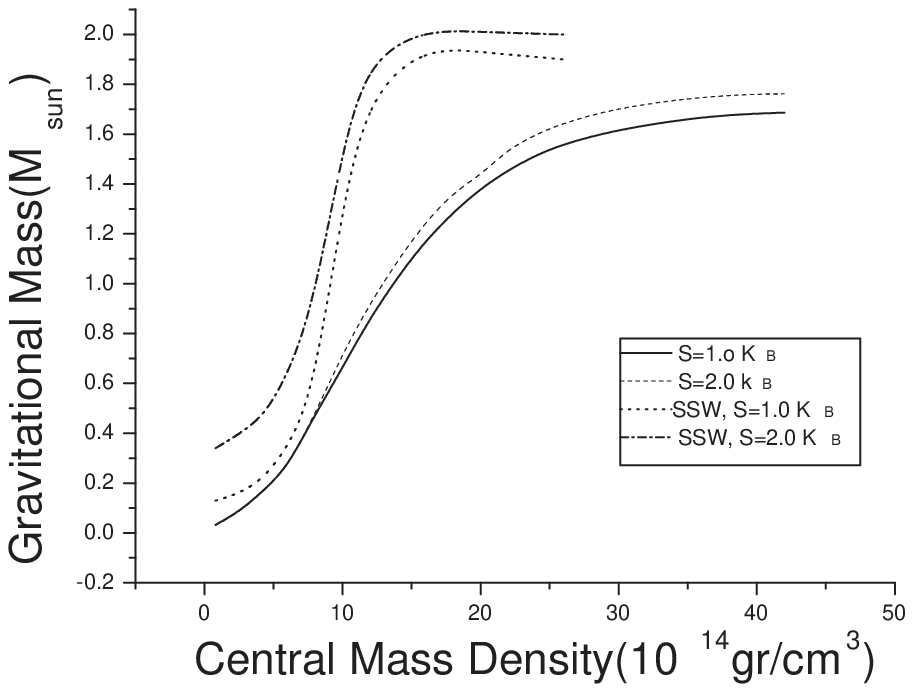}
 \caption{As Fig. 6, but for the gravitational mass ($M_\odot$) of
 beta-stable protoneutron star versus central mass density ($10^{14}
 gr/cm^3$).
   }
  \label{fig7}
\end{figure}
\newpage
\begin{figure}
 \includegraphics{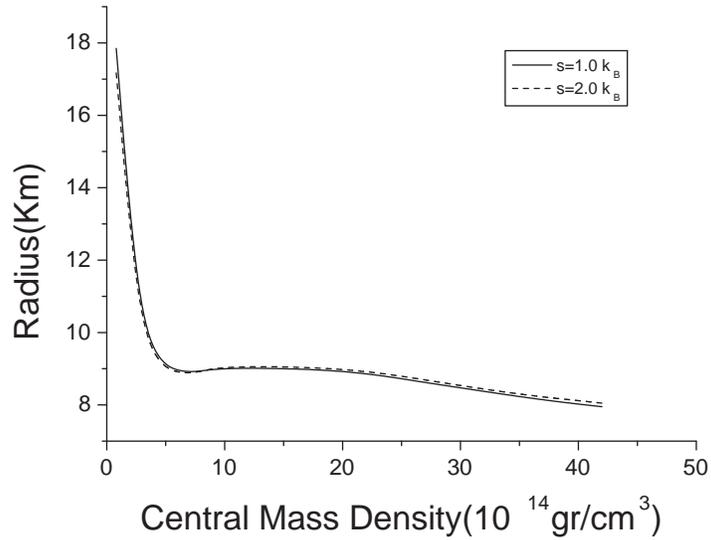}
 \caption{Our results for the radius ($km$) of
 beta-stable protoneutron star versus central mass density ($10^{14} gr/cm^3$)
 at different entropies.
  }
  \label{fig8}
\end{figure}
\newpage
\begin{figure}
 \includegraphics{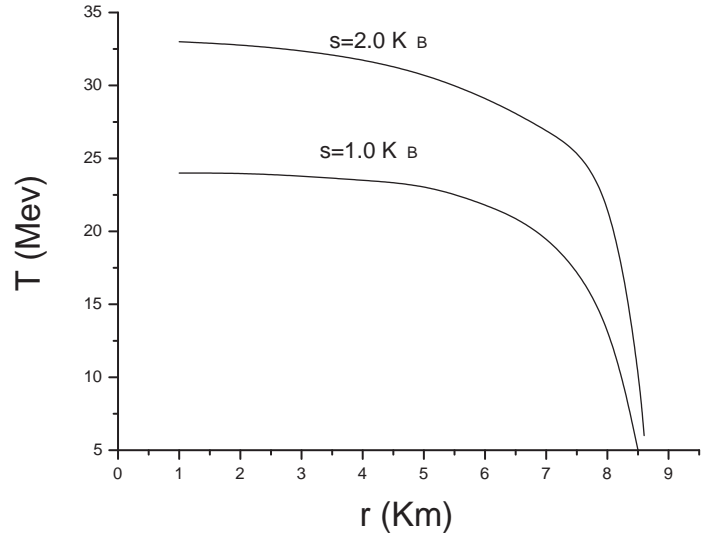}
 \caption{Our results for the temperature profile  of
 beta-stable protoneutron star at different entropies.
  }
  \label{fig9}
  \end{figure}
\newpage
\begin{figure}
 \includegraphics{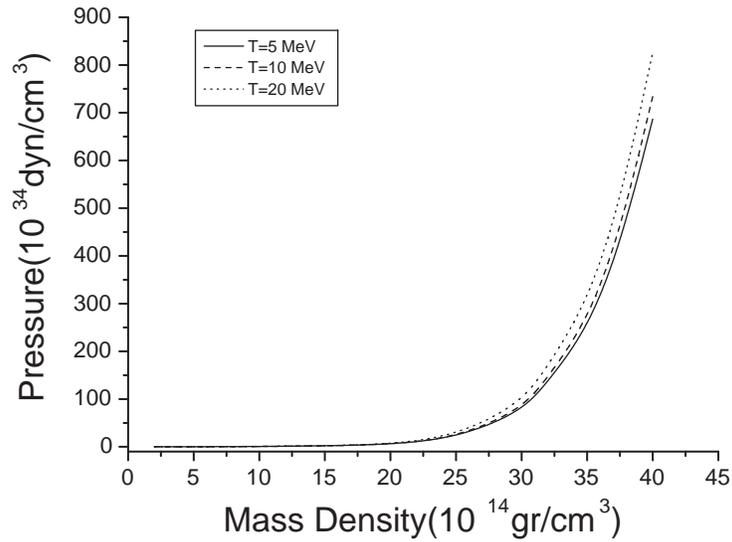}
 \caption{Our results for the pressure ($10^{34} dyn/cm^2$) of
 beta-stable protoneutron star matter as a
 function of mass density ($10^{14} gr/cm^3$) at different temperatures.
  }
  \label{fig10}
\end{figure}

\newpage
\begin{figure}
 \includegraphics{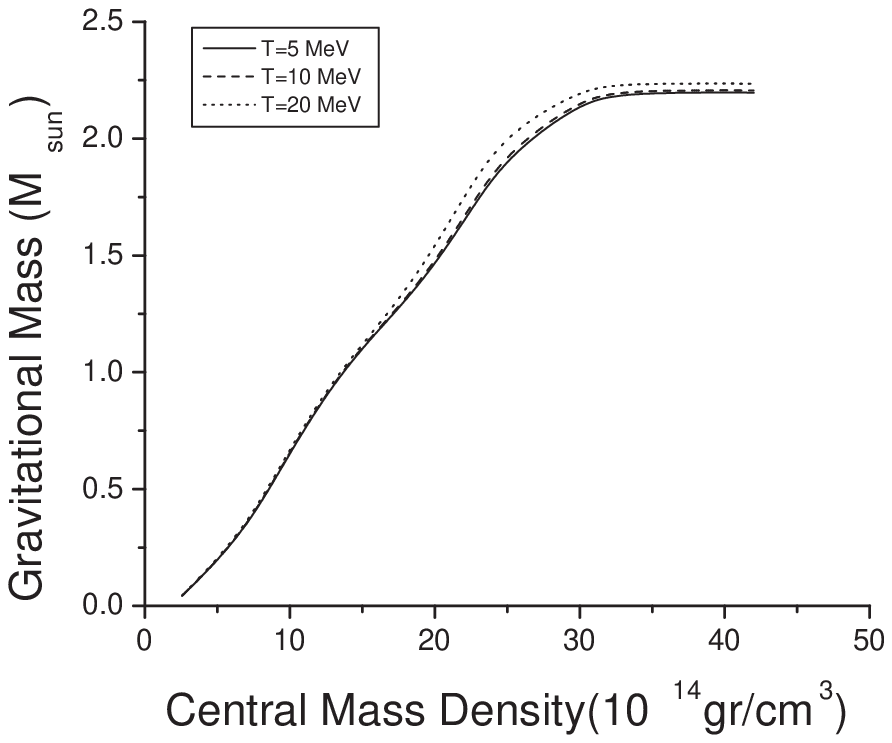}
 \caption{As Fig. 10, but for the gravitational mass ($M_\odot$)
of
 beta-stable protoneutron star versus central mass density ($10^{14}
 gr/cm^3$).
   }
  \label{fig11}
\end{figure}
\newpage
\begin{figure}
 \includegraphics{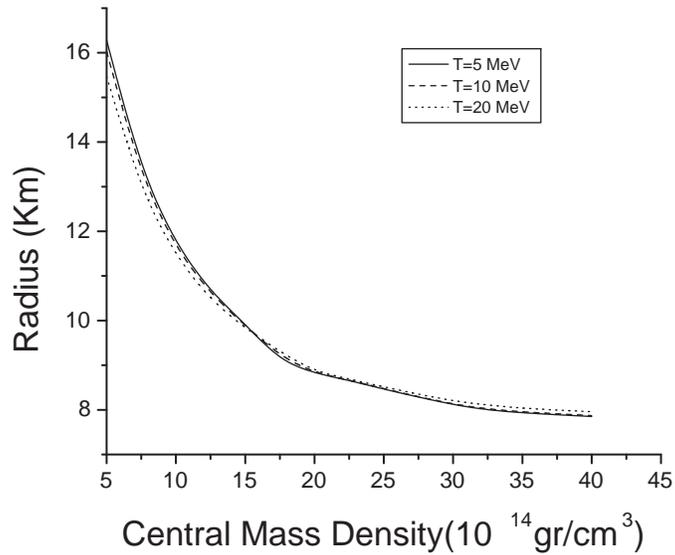}
 \caption{As Fig. 10, but for the radius ($km$) of
 beta-stable protoneutron star versus central mass density ($10^{14}
 gr/cm^3$).
    }
  \label{fig12}
\end{figure}

\end{document}